\documentclass[12pt]{article}
 \usepackage{graphicx}
 \usepackage{amssymb,amsmath}

\begin{document}

\begin{center}
{\Large\bf \boldmath Collisional Stark Transitions and Induced
Annihilation of Cold Antiprotonic Helium Ions}

\vspace*{6mm}
 {G.Ya. Korenman and S.N. Yudin}\\
 {\small \it D.V.Skobeltsyn Institute of Nuclear Physics,
 M.V. Lomonosov Moscow State University,\\
Moscow 119991,  Russia }

E-mail: korenman@nucl-th.sinp.msu.ru

\end{center}

\vspace*{6mm}

\begin{abstract}
Stark transitions and induced annihilation of antiproton in the
collisions of antiprotonic helium ions
$(\bar{p}\mathrm{He}^{+2})_{nl}$ with $\mathrm{He}$ atoms at very
low energy ($\sim 10$ K) are considered in the framework of
quantum coupled-channels method taking into account all the states
with different $l$ at given $n\sim 30$, including the annihilating
$ns$ and $np$-states. Elastic scattering, Stark transitions and
induced annihilation during collisions are produced by scalar and
dipole terms in the interaction. It is shown that the most
important contribution to the processes comes from the long-range
polarization interaction. Admixtures of the $ns$ and $np$-states
to the states with higher $l$ during collisions induce the
effective annihilation cross sections for the initial $l$ up to
15, but don't affect the Stark cross sections for the initial
states nearly to circular orbits. Total rates of the Stark
transitions from the circular orbits with $n=28 - 32$, averaged
over the thermal motion, are compatible with the recent ASACUSA
data. Isotope effect as well as the dependence on $n$ are also
qualitatively agree with the experiment.
\end{abstract}

\vspace*{6mm}

\section{Introduction}
The experimental discovery and investigation of the metastable
antiprotonic states in helium have opened a new chapter in the
study of antiprotonic atoms (see a comprehensive review
\cite{ref1} of the results obtained up until 2001). In addition to
the fundamental properties of the antiproton, these investigations
give interesting insights into the interaction of antiprotonic
helium with ordinary atoms and molecules. These data stimulated
many theoretical papers on the mechanisms of antiprotonic helium
formation, on collisional quenching of the metastable states, on
effects of collisional (density) shift and broadening of
E1-spectral lines, {\it etc}. New stage of the study of
antiprotonic helium by the ASACUSA collaboration at the AD beam
(CERN) brings further new data that pose specific problems in the
theory of collisional effects at low temperature ($\sim 10$ K).
Thus for the measurements of hyperfine splitting of antiprotonic
helium levels by a triple laser-microwave-laser resonance method
\cite{ref2}, it is important to estimate the rates of collisional
transitions between the HFS sublevels, and the density shifts and
broadenings of the microwave M1 spectral lines \cite{ref3}. The
first observation of cold long-lived antiprotonic helium ions
$(\bar{p}\mathrm{He}^{2+})$ with lifetimes $\tau_i\sim 100$ ns
against annihilation \cite{ref4} revealed an unexpected
$n$-dependence and isotope effect for effective annihilation rates
in circular states ($l=n-1$) at $n=28 - 32$. Earlier
considerations of collisional processes of
$(\bar{p}\mathrm{He}^{2+})$ at higher energy could not explain
these data.

In this paper we consider Stark transitions and induced
annihilation in the collisions of antiprotonic helium ion with
$\mathrm{He}$ atoms at very low energy ($\sim 10$ K),
\begin{equation} \label{eq1}
(\bar{p}\mathrm{He}^{+2})_{nl} +\mathrm{He}  \rightarrow
\begin{cases} (\bar{p}\mathrm{He}^{+2})_{nl'} +\mathrm{He} \quad
\text{(Stark)} ,\\
 (\pi\ldots \pi X)+ \mathrm{He} \quad
\text{(annihilation)}.
\end{cases}
\end{equation}
We use the quantum close-coupling method with the appropriate
interatomic interaction. The boundary conditions at
$R\rightarrow\infty$ in the channels corresponding to $ns$ and
$np$-states of the antiprotonic ion are modified with regard to
the annihilation widths of the states.

\section{Interaction between antiprotonic helium ion and
$\mathrm{He}$ atom}

Let us discuss the interaction between a
$(\bar{p}\mathrm{He}^{+2})_{nl}$ ion and an ordinary $\mathrm{He}$
atom at low energy ($E<E_a=27.2$ eV). The relative velocity of the
collision is small as compared with the electron velocity in the
atom, $v \ll v_e$, where $v_e\sim Z_{eff} v_a$ is a characteristic
electron velocity and $v_a=e^2/\hbar$ is the atomic unit of
velocity. (Hereafter we use the atomic system of units
$\hbar=e=m_e=1$, except as otherwise indicated explicitly.) The
orbital velocity of the antiproton in the states with the
principal quantum number $n\gg 1$ is also small, $v_n=Z/n \ll 1$.
At these conditions the heavy particles ($\bar{p}$ and nuclei) in
the system are slow as compared with electrons, therefore the
electronic variables can be separated out within adiabatic
approximation. Then the sum of Coulomb interactions between the
antiprotonic ion $(\bar{p}\mathrm{He}^{2+})$ and $\mathrm{He}$
atom is replaced by the effective three-body interaction
$V(\mathbf{R,\, r})$ that depends on the relative coordinates
$\mathbf{R}$ of two colliding subsystems and on the inner
coordinates $\mathbf{r}$ of the antiprotonic ion.

At a large distance between two subsystems the interaction is
reduced to the long-range polarization term, $V(\mathbf{R,\,
r})\rightarrow -\alpha/2R^4$, where $\alpha=1.383$ a.u. is a
polarizability of the $\mathrm{He}$ atom. It is well known that
the classical particle in a such potential can fall to the center,
that is treated as a 'polarization capture' with the effective
cross section
\begin{equation} \label{eq2}
\sigma_p(E)=\pi\sqrt{2\alpha/E}.
\end{equation}
Then a mean radius of the effective interaction can be estimated
as $\bar{R} \sim (\alpha/E)^{1/4}$, i.e., at least several atomic
units at $E<1$ eV. On the other hand, a mean size of the
antiprotonic helium ion can be estimated as $\bar{r}=n^2/(\mu Z)$,
where $\mu$ is the reduced mass of the $\bar{p} -\mathrm{^A
He}^{2+}$ system, $\mu \simeq 1467 m_e$ for A=4  and 1376$m_e$ for
A=3. At $n= 30$ we have $\bar{r}\simeq 0.3 a_0 \ll \bar{R}$.
Therefore the three-body potential $V(\mathbf{R,\, r})$ can be
developed in the multipolar series
\begin{equation} \label{eq3}
 V(\mathbf{R,r}) \simeq V_0(R) + (\mathbf{d}\cdot\nabla_{\mathbf{R}})V_0(R) +
 \left( Q_2(\mathbf{r})\cdot C_2(\Omega_R) \right) V_0^{\prime\prime}(R)  ,
 \end{equation}
where $\mathbf{d}=-\xi_1\mathbf{r}$ and $Q_{2\nu}=- \xi_2 r^2
C_{2\nu}(\Omega_r)$ are dipole and quadruple operators of the ion,
$C_{2\nu}(\Omega)=Y_{2\nu}(\Omega)\sqrt{4\pi/5}$. The factors
$\xi_1 =(m_A+2m_p)/(m_A+m_p)$ and
$\xi_2=(m_A^2-2m_p^2)/(m_A+m_p)^2$ arise owing to a shift of the
center of charge from the center of mass. The scalar term $V_0(R)$
in \eqref{eq3} can produce only the elastic scattering, whereas
the dipole and quadruple terms can mix the $nl$ states of the ion
and cause the Stark transitions $(nl\rightarrow nl')$ of the
antiprotonic helium ion.

The potential $V_0(R)$ is an adiabatic potential of the
interaction between the ionic charge (+1) and He atom, that is, in
fact, the same as the adiabatic potential of the interaction
between proton and He atom. The latter, according to the
calculations \cite{ref5}, can be approximated by a sum  of the
Morse potential $V_M(R)$ and of the polarization long-range
interaction $V_p(R)$ cut off at an intermediate distance,
\begin{align}
V_0(R) & = V_M(R)+ V_p(R), \label{eq4} \\
V_M(R) & = D_0 \left\{\exp[-2\beta(R-R_e)] -
            2\exp[-\beta(R-R_e)]\right\}, \label{eq5} \\
V_p(R) & = -\frac{\alpha}{2R^4} \left\{1 -
\exp\left[-\gamma(R-R_e)^4\right]\right\}\theta(R-R_e) ,
\label{eq6}
\end{align}
For the calculations we take the following values of the
parameters \cite{ref5,ref6} $D_0=0.075$, $R_e =1.46$, $\beta
=1.65$, $\gamma=0.005$ a.u. It should be noted that the Morse
potential \eqref{eq5} contains a strong repulsion at $R<R_w=R_e-
\beta^{-1} \ln 2$. In order to avoid the loss of accuracy at
$R<R_w$ in the solving of the coupled-channels equations at very
low energy, we introduce the infinite wall at $R_w=1.0399$ a.u.

\section{Coupled-channels equations for the collisions
$(\bar{p}\mathrm{He}^{2+})$ ions with $\mathrm{He}$ atoms}
 In the previous section we noted that electronic variables in the system
$(\bar{p}\mathrm{He}^{+2})_{nl} +\mathrm{He}$ can separated out
within the adiabatic approximation. Then total effective
hamiltonian of the three heavy particles $(\bar{p}-Z-\mathrm{He})$
is
 \begin{equation} \label{eq7}
  H = T(\mathbf{R}) + h(\mathbf{r}) + V(\mathbf{R,\, r}) ,
\end{equation}
where $T(\mathbf{R}) = (-1/2m)\nabla^2_{\mathbf{R}}$ is the
kinetic energy operator, $m$ is the reduced mass of the colliding
subsystems, and $h(\mathbf{r})$ is the inner hamiltonian of the
antiprotonic ion. For our aims it can be presented as
\begin{equation} \label{eq8}
h =h_0 +U_{\mathrm{opt}}(r),
\end{equation}
where $h_0$ is the hamiltonian of hydrogen-like atom with the
reduced mass $\mu$ and with the nuclear charge $Z=2$, and
$U_{\mathrm{opt}}(r)$ is a short-range complex optical potential
with the absorptive imaginary part
$\mathrm{Im}U_{\mathrm{opt}}(r)\leq 0$. Let $E_{nl}$ and
$\phi_{nlm}(\mathbf{r})=R_{nl}(r)Y_{lm}(\Omega_r)$ be the
eigenvalues and the eigenfunctions of the hamiltonian \eqref{eq8},
corresponding to the discrete levels of the
$(\bar{p}\mathrm{He}^{2+})$ system,
\begin{equation} \label{eq9}
h \phi_{nlm}(\mathbf{r}) = E_{nl}\phi_{nlm}(\mathbf{r}).
\end{equation}
In addition, for the conjugated hamiltonian $h^+$ we introduce
eigenvalues $\widetilde{E}_{nl}= E_{nl}^*$ and eigenfunctions
$\widetilde{\phi}_{nlm}(\mathbf{r})=\widetilde{R}_{nl}(r)Y_{lm}(\Omega_r)$
with $\widetilde{R}_{nl}(r)=R^*_{nl}(r)$. Two sets of the
eigenfunctions, $\phi_{nlm}(\mathbf{r})$ and
$\widetilde{\phi}_{nlm}(\mathbf{r})$, form jointly a bi-orthogonal
system. The eigenvalue $E_{nl}$ in general case contains a
negative imaginary part,
 $E_{nl}=E_{nl}^R - \mathrm{i}\Gamma_{nl}/2$,
where $\Gamma_{nl}$ is the annihilation width of the level. The
real part of the eigenvalue is usually expressed as $E_{nl}^R=e_n
- \epsilon_{nl}$, where $e_n = -\mu Z^2/2n^2$ is the energy of the
hydrogen-like system with the point nuclear charge, and
$(-\epsilon_{nl})$ is the energy shift of the level due to the
nuclear interaction and the finite distribution of the nuclear
charge.  Total difference $(E_{nl} - e_n)$ is usually referred to
as the complex energy shift $\Delta E_{nl}=-\epsilon_{nl} -
\mathrm{i}\Gamma_{nl}/2$. The states with $l\geqslant 2$ of the
antiprotonic helium ion can be considered as hydrogen-like,
however $ns$ and $np$-states have considerable complex energy
shifts. Dependence of $\Delta E_{nl}$ on $n$ is well known from
the theory, $\Delta E_{ns} = \Delta E_{1s}/n^3$,
 $\Delta E_{np}=32(n^2-1)/(3n^5)\cdot \Delta E_{2p}$. The
remaining parameters are taken as a combination of the
experimental data and theoretical calculations using optical
model. For $^4$He we use $\Gamma_{1s}=11$ keV, $\Gamma_{2p}=36$
eV, $\epsilon_{nl}=0.3 \Gamma_{nl}$.

Let us introduce the basis function
 \begin{equation} \label{eq11}
\Phi_k^{JM\pi}(\mathbf{r},\Omega_R)=
\left(\phi_{nl}(\mathbf{r})\otimes Y_L(\Omega_R)\right)_{JM} ,
\end{equation}
where $L$ is the quantum number of the relative angular momentum
of the colliding subsystems, $\pi=(-1)^{l+L}$ is the parity of the
system, and $J,M$ are the quantum numbers of the total angular
momentum $(\mathbf{J=l+L})$. Multi-index $k$ involves the set of
the quantum numbers $\{n,l,L\}$. In addition, we introduce also
the functions $\widetilde{\Phi}_k^{JM\pi}(\mathbf{r},\Omega_R)$,
which contain $\widetilde{R}_{nl}(r)$ and form with \eqref{eq11} a
bi-orthogonal system. Total wave function of the system
$(\bar{p}\mathrm{He}^{2+})+\mathrm{He}$ in the state with the
total energy $E$, the total angular momentum $J$ and the parity
$\pi$ in the framework of the close-coupling approximation can be
written as
\begin{equation} \label{eq12}
 \Psi_i^{EJM\pi}(\mathbf{R,r}) =\sum_k \Phi_k^{JM\pi}(\mathbf{r},\Omega_R) \psi_{ki}^{EJ\pi}(R)/R,
 \end{equation}
where $i$ is the index of the incoming channel. Substituting
\eqref{eq12} into the Schroedinger equation with the effective
hamiltonian \eqref{eq7} and taking into account the
bi-orthogonality property of the basis states, we get the system
of coupled-channel equations
\begin{equation} \label{eq13}
 \psi_{ji}^{\prime\prime}(R) +  \left[k_j^2- L_j(L_j+1)/R^2\right] \psi_{ji}(R)
= 2m\sum_k V_{jk}(R)\psi_{ki}(R),
\end{equation}
where $k_j^2=2m(E-E_{nl})$ and $ V_{jk}(R)= \langle
\widetilde{\Phi}_j |V(\mathbf{R,\, r})| \Phi_k \rangle $. For the
channels with non-zero $\Gamma_{nl}$ we assume $\mathrm{Im}k_j>0$.
The conserved quantum numbers $(E,J,M,\pi)$ are omitted the for
the sake of brevity.

The system \eqref{eq13} looks very similar to the commonly used in
the standard close-coupling methods. However in our problem the
wave numbers $k_j$ in some channels contain imaginary parts,
therefore the conventional boundary conditions at
$R\rightarrow\infty$ (incoming + outgoing waves) in these channels
have no sense and should be modified. (Of course, the boundary
conditions at the origin $\psi_{ji}(0)=0$ remain valid.) For
simplicity, we will not consider the closed channels with
$E_{nl}^R>E$. Let $\alpha$ be the subspace of the normal channels
(with $\Gamma_{nl}=0$) and $\beta$ the subspace of the
annihilating channels with $\Gamma_{nl}\neq 0$ and
$\mathrm{Im}k_j>0$. The channels from the subspace $\beta$ can not
be presented among the incoming channels, therefore the total wave
function $\Psi_i^{EJM\pi}(\mathbf{R,r})$ at $i\in \beta$ has to be
zero everywhere in $R\in[0,\infty)$, i.e.,
\begin{equation} \label{eq14}
\psi_{ji}(R) \equiv 0 \text{ at } i\in \beta, \text{ any } j.
\end{equation}
For the ordinary incoming channels $(i\in\alpha)$ the components
$\psi_{ji}(R)$ with $j\in\alpha$ can contain the both incoming (at
$j=i$) and outgoing waves, whereas the components with $j\in\beta$
have to damp out at the large distance. They also can be presented
in the form of 'outgoing' waves with the complex $k_j$
($\mathrm{Im} k_j \geqslant 0$). Hence the asymptotic boundary
conditions at $R\rightarrow\infty$ for the functions
$\psi_{ji}(R)$ at $i\in\alpha$ and any $j$ can be written as
 \begin{equation} \label{eq15}
\psi_{ji}(R) \rightarrow k_i^{-1/2} \exp[-\mathrm{i}(k_i
R-L_i\pi/2)] \delta_{ij} - k_j^{-1/2}\exp[\mathrm{i}(k_j
R-L_j\pi/2)] C_{ji}  \text{~ at }  i\in\alpha,
\end{equation}
For the channels $j\in\alpha$ this asymptotic expression has the
traditional sense, and the coefficients at the outgoing waves give
the elements of $S$-matrix, $S_{ji}= C_{ji}$, corresponding to the
transitions between the channels in the subspace $\alpha$. For the
channels $j\in\beta$ the first term in \eqref{eq13} is absent
because of $j\neq i$, whereas the second term is, in fact, damped
out:
 \begin{equation} \label{eq16}
\psi_{ji}(R) \sim \exp(-\mathrm{Im}k_j\cdot R) C_{ji}
      \text{~ at }  i \in\alpha, j \in\beta.
 \end{equation}
The density current of the outgoing  waves in these channels is
also damped out exponentially, therefore the coefficients $C_{ji}$
at $j\in\beta, i\in\alpha$ can not be treated as the $S$-matrix
elements of the transitions $(\alpha\to\beta)$.

The $S$-matrix in the subspace $\alpha$ gives elastic and
inelastic cross sections
\begin{equation} \label{eq17}
\sigma(nl\rightarrow n'l') = \frac{\pi}{k_i^2} \sum_{L,L',J}
\frac{2J+1}{2l+1} \left|\langle n'l'L'|S^J|nlL\rangle
-\delta_{nn'}\delta_{ll'}\delta_{LL'}\right|^2 .
\end{equation}
Due to absorptive part of the hamiltonian $h$ the $S$-matrix is
non-unitary. The defect of unitary  gives the cross section of
induced annihilation for the definite initial state $nl$,
\begin{equation} \label{eq18}
\sigma_{nl}^a = \frac{\pi}{k_i^2} \sum_{L,J}
\frac{2J+1}{(2l+1)(2L+1)} \left[1 -\sum_{n'l'L'} |\langle
n'l'L'|S^J|nlL\rangle |^2 \right].
\end{equation}

More detailed discussion of the problem of boundary conditions for
the close-coupling method with annihilating states included in the
basis, as well as the scheme of the numerical solution of the
system \eqref{eq13} satisfying to the boundary conditions
$\psi_{ji}(0)=0$ at the origin, to the constraint \eqref{eq14} at
$i\in\beta$ and to the boundary conditions \eqref{eq15} at
$R\rightarrow \infty$, will be given in the separate paper
\cite{ref7}.

\section{Stark transitions and induced annihilation
in antiprotonic helium ions}

With interaction \eqref{eq3} and complex energy shifts of $ns$ and
$np$-states we have solved the coupled-channel equations using the
basis that includes all states with different $l$ at fixed $n$ and
relative angular momenta up to $L=12$ at $E\sim 10$ K. For the
calculation of matrix elements $V_{jk}(R)$ we use the additional
approximation $R_{nl}(r)=\widetilde{R}_{nl}(r)
=R^{H-like}_{nl}(r)$ for all states included the basis.

Cross sections of elastic scattering, of partial and total Stark
transitions and of induced annihilation for the initial states
with $l\geqslant 2$ were obtained. A relative importance of the
different terms in the interaction can be illustrated by the total
cross sections of the Stark transitions from the circular state
with $n=30$ at $E=10$ K. If we take the short-range Morse
potential for $V_0(R)$,  the cross section is about 120 a.u. When
the polarization term $V_p(R)$ is included into $V_0(R)$, the same
cross section rises to 620 a.u., whereas the quadruple term in
\eqref{eq3} changes the cross section by the negligible value
($<0.2\%$). Therefore the main part of the calculations was done
without the quadruple interaction.

\begin{figure}[ht] \vspace*{-5mm}
\begin{minipage}{16pc}
\includegraphics[width=16pc,height=100mm]{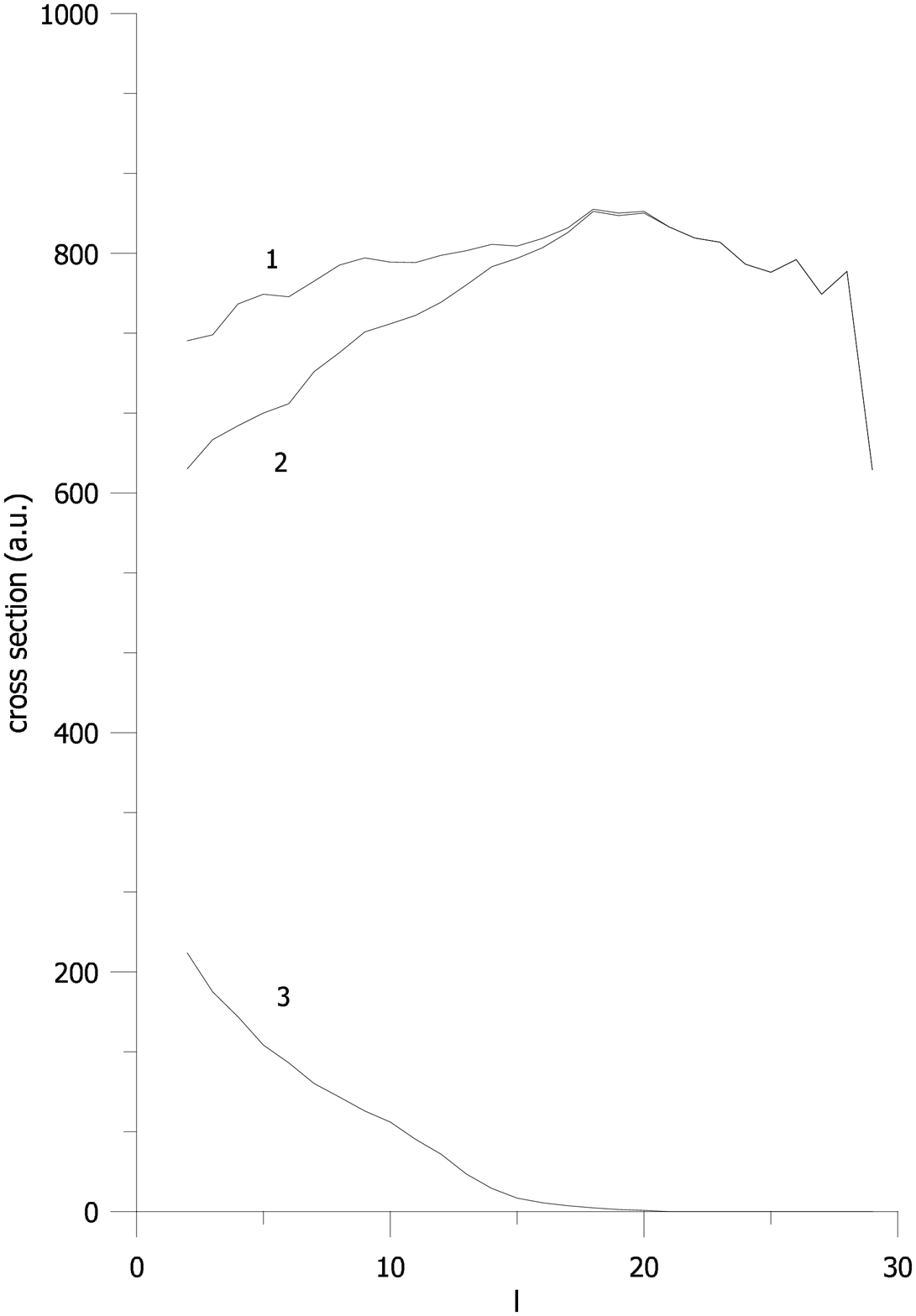}
\caption{\label{sigma_vsl}Dependence of total Stark (curves 1 and
2) and induced annihilation (curve 3) cross sections on initial
$l$ ($n =30,\, E =10$ K). The curve 1 is obtained without
including the annihilation widths.}
\end{minipage}\hspace{2pc}%
\begin{minipage}{18pc}
\includegraphics[width=18pc,height=100mm]{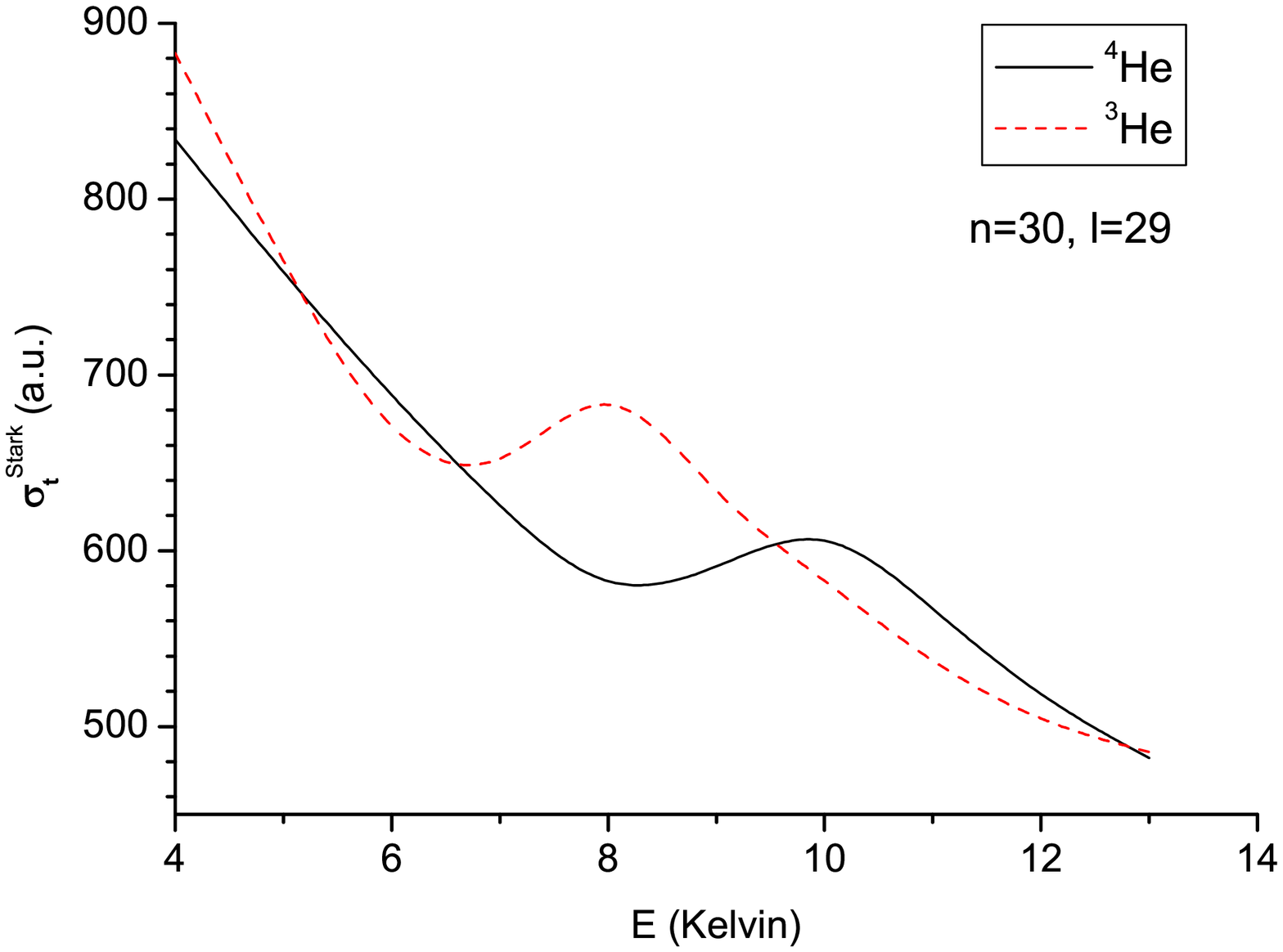}
\caption{\label{sigma_vsE} Energy dependence of the Stark cross
section for the circular orbit with $n=30$ for $^4$He (full curve)
and $^3$He (dashed curve).}
\end{minipage}
\end{figure}

Figure \ref{sigma_vsl} shows the dependence of total Stark and
induced annihilation cross sections on the initial quantum number
$l$ of the antiprotonic ion at $n=30$ and $E=10$ K. The induced
annihilation has an appreciable cross sections for $l\lesssim 15$.
At higher initial $l$, this process is negligible and doesn't
affect the Stark transitions. It should be noted that the total
Stark cross sections at these energies are very large, of several
hundreds atomic units. They are comparable with the cross section
\eqref{eq2} of 'polarization capture' ($\sigma_p\simeq 930$ a.u.
at $E=10$ K). Fig. \ref{sigma_vsE} shows the energy dependence of
the total Stark cross sections for circular orbits with $n=30$ for
two isotopes $^4$He and $^3$He. The curves look similar, but don't
give simple isotopic scaling. Dependencies of the Stark cross
sections on the principal quantum number $n$ for two isotopes at
the c.m. energy $E=10$ K are illustrated by Table \ref{tab1}.
\begin{table}[ht]
\caption{\label{tab1} Total Stark cross sections (in a.u.)  for
the circular orbits at c.m. energy $E=10$ K depending on the
principal quantum number $n$ for two isotopes.}
\begin{center}
\begin{tabular}{c|c|c|c|c|c} \hline
       &\multicolumn{5}{c}{$n$} \\ \cline{2-6}
 isotope    &28    &29       &30     &31     &32  \\ \hline
 $^4$He     &573.5 &598.2    & 619.6 &636.5  &649.0 \\
 $^3$He     &539.3 &566.5    &583.4  &597.0  &605.1\\ \hline
\end{tabular}
\end{center}
\end{table}

\begin{figure}[ht] \vspace*{-8mm}
\includegraphics[width=150mm, height=125mm]{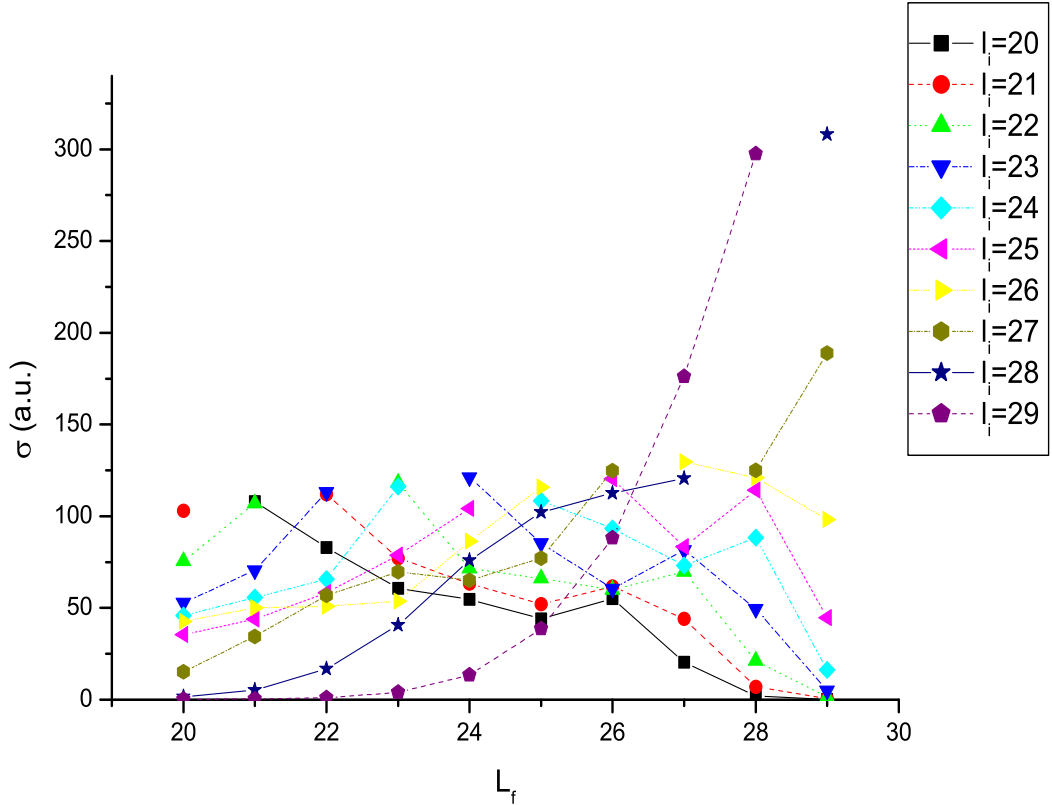}
\vspace*{-8mm}
 \caption{\label{sigma_lilf} Partial Stark cross
sections $\sigma(nl_i\rightarrow nl_f)$ for collisions
$(\bar{p}^4\mathrm{He}^{2+})_{nl} + ^4\mathrm{He}$, at $n=30$,
$E=10$ K.}
\end{figure}

Partial Stark cross section $\sigma(nl_i\rightarrow nl_f)$ for
$^4$He, $n=30$, at $E=10$ K are shown on Fig. \ref{sigma_lilf}.
(Elastic cross sections $\Delta l=0$ are not shown here, because
they are out of the boundary of the figure.) It is seen from the
figure that the cross sections tend to diminish with $\Delta
l=|l_i-l_f|$. Nevertheless, in spite of the fact that the coupling
between channels is carried out by the dipole interaction, the
cross sections remain appreciable up to $\Delta l$ of several
units due to multi-step transitions that are included in the
close-coupling equations.

\begin{table}[ht]
\caption{\label{kappa} Per-atom Stark transition rates at $T=10$ K
for circular orbits of two helium isotopes (units
$10^{-16}\,\mathrm{MHz\cdot cm^3}$).}
\begin{center}
\begin{tabular}{llll} \hline
isotope& $n=28$& $n=30$& $n=32$ \\ \hline
$^4$He & 3.77  & 4.15  & 4.46   \\
$^3$He & 4.46  & 4.82  & 5.01   \\ \hline
\end{tabular}
\end{center}
\end{table}

Per-atom rates of Stark transitions from circular orbits,
$\kappa_{St}=\langle v\sigma_{St} \rangle$, averaged over the
thermal motion at $T=10$ K, are given in Table \ref{kappa} for the
two isotopes. The calculated values can not be compared directly
with the experimental values \cite{ref3} of per-atom effective
annihilation rates $d\gamma/d\rho \sim (1 - 3)\cdot
10^{-16}\,\mathrm{MHz\cdot cm^3}$, because the latter rates are a
result of many cascade processes. Nevertheless, Stark transitions
from circular orbits can be considered as the most important first
step of the cascade, because they change the inner angular
momentum of the antiprotonic ion and open up possibilities for
other processes. It is important that the values of $\kappa_{St}$
are greater than (but of the same order of value as) the
experimental rates $d\gamma/d\rho$. Moreover, we obtain also that
$\kappa_{St}$ for $^3$He is greater than for $^4$He and grows with
$n$, and these tendencies are correlated with the observed
dependencies of the value $d\gamma/d\rho$.

 \section{Conclusion}

We have considered Stark transitions and induced annihilation in
the collisions of cold $(\bar{p}\mathrm{He}^{2+})_{nl}$ ions with
$\mathrm{He}$ atoms, using the coupled-channels method with the
annihilating states ($ns$ and $np$) included in the basis and with
the respectively modified boundary conditions. The results of the
calculations for the states with $n\sim 30$ show that the induced
annihilation is appreciable for the initial states with orbital
angular momenta $l\lesssim 10 - 15$. For the larger $l$, the
induced annihilation is negligible and the Stark cross sections
are not affected by the annihilating states. Total Stark cross
sections are comparable by order of value with the classical
polarization capture cross section. Partial Stark cross sections
are appreciable for the transitions with $\Delta l$ up to several
units.

Per-atom rates of Stark transitions $\kappa_{St}$ from circular
orbits for two helium isotopes, averaged over the thermal motion,
reveal a some increase of the rates with $n$ and the isotope
effect, which show the (12 - 18)\% excess in the rates for
$^3\mathrm{He}$. These effects are correlated with the observed
dependencies of the effective annihilation rates $d\gamma/d\rho$
for the initial circular orbits.  The calculated and measured
values, $\kappa_{St}$ and $d\gamma/d\rho$, respectively, are
different but closely coupled in the physical content. We have
obtained also that the values of $\kappa_{St}$ are greater than,
but of the same order of value as the experimental rates
$d\gamma/d\rho$. Therefore we conclude that our results are
compatible with the experimental data on the collisional
properties of cold antiprotonic helium ion in the medium.

\section*{Acknowledgments}
This work was supported by Russian Foundation for Basic Research
by the grant 06-02-17156. One of the authors (G.K.) thanks to
R.~Hayano and E.~Widmann for the useful discussions.

\end{document}